\documentclass[journal,10pt,compsoc,twoside]{IEEEtran}

\usepackage{amsmath}
\usepackage{amssymb}
\usepackage{amsfonts}
\usepackage{booktabs}
\usepackage{graphicx}
\usepackage{cite}
\usepackage{url}
\usepackage{listings}
\usepackage{xcolor}
\usepackage{microtype}
\usepackage{bm}
\usepackage{multirow}
\usepackage{array}
\usepackage{tabularx}

\definecolor{codegreen}{rgb}{0,0.6,0}
\definecolor{codegray}{rgb}{0.5,0.5,0.5}
\definecolor{codepurple}{rgb}{0.58,0,0.82}
\definecolor{backcolour}{rgb}{0.95,0.95,0.92}

\lstdefinestyle{pythonstyle}{
    backgroundcolor=\color{backcolour},
    commentstyle=\color{codegreen},
    keywordstyle=\color{magenta},
    numberstyle=\tiny\color{codegray},
    stringstyle=\color{codepurple},
    basicstyle=\ttfamily\footnotesize,
    breakatwhitespace=false,
    breaklines=true,
    captionpos=b,
    keepspaces=true,
    numbers=left,
    numbersep=5pt,
    showspaces=false,
    showstringspaces=false,
    showtabs=false,
    tabsize=2
}

\begin{document}

\title{When Does Few-Shot Prompting Help?\\
A Systematic Empirical Study of Shot-Count Effects\\
Across Model Scale, Architecture, and Output Parsing Robustness}

\author{Ayush~Dwivedi~and~Ashvi~Soni%
\thanks{Code:
\protect\url{https://github.com/a-dwivedi/few-shot-prompting-study}.
Manuscript submitted as a preprint to arXiv, July 2026.}}

\markboth{PREPRINT --- SUBMITTED TO ARXIV, JULY 2026}%
{Dwivedi and Soni: When Does Few-Shot Prompting Help?}

\maketitle

\begin{abstract}
Few-shot prompting, defined as the practice of prepending a small number of
input-output demonstration pairs to a query before presenting it to a
large language model (LLM), is among the most widely adopted
inference-time techniques in applied natural language processing.
Yet surprisingly little systematic empirical work investigates how
shot count interacts with model scale, architecture, and output format
compliance in determining classification performance.
This paper presents a controlled experimental study of five LLMs across
six shot-count configurations ($k\in\{0,1,2,3,5,8\}$) on the AG News
four-class text classification benchmark ($n=200$, seed fixed for
reproducibility). Our models span proprietary and open-source families:
Gemini Flash Lite (Google), GPT-4o-mini (OpenAI), Llama~3.1~8B Instant,
Llama~3.3~70B Versatile, and Llama~4~Scout~17B (all free-tier
open-source models hosted on Groq). We report macro-averaged F1 scores
with 95\% bootstrap confidence intervals ($B=10{,}000$),
permutation-test $p$-values, and Cohen's~$d$ effect sizes across all
30 model--shot configurations. Our findings reveal four qualitatively
distinct behavioral regimes: (1)~models already well-calibrated at
zero-shot that show modest, statistically insignificant gains from
demonstrations (Gemini, GPT-4o-mini); (2)~models that undergo
catastrophic zero-shot failure but recover dramatically with a single
example (Llama~3.1~8B, Cohen's $d=10.98$, $p<0.0001$); (3)~models
that are optimal at zero-shot and degrade monotonically with additional
examples (Llama~4~Scout); and (4)~models exhibiting a U-shaped
performance curve (strong at zero-shot, sharply degraded at low
shot counts, and partially recovering at higher counts), independent of
output parsing issues (Llama~3.3~70B: 0-shot
$\text{F1}=0.907\rightarrow$ 2-shot $\text{F1}=0.635\rightarrow$
5-shot $\text{F1}=0.785$ with parser corrected). We additionally
identify, diagnose, and correct a systematic parsing artifact that
artificially deflated Llama~3.3~70B performance by up to 206\%,
constituting a methodological contribution to LLM evaluation practice.
Our results demonstrate that the relationship between shot count and
classification performance is not monotonic, not universal, and not
predictable from model scale alone, with significant implications
for practitioners deploying LLMs in production classification pipelines.
\end{abstract}


\IEEEpeerreviewmaketitle

\section{Introduction}

\IEEEPARstart{T}{he} emergence of large language models capable of
performing complex reasoning tasks from natural language instructions
has precipitated a fundamental shift in how practitioners approach
supervised classification. Where previously one might fine-tune a
BERT-class encoder on labeled data, the contemporary default is
increasingly to query a pre-trained LLM with carefully constructed
prompts, sometimes augmented with a handful of labeled examples drawn
from the task distribution. This paradigm, known as \emph{few-shot
prompting} or \emph{few-shot in-context learning} (ICL), was
popularized by Brown et al.~\cite{brown2020language} with GPT-3 and
has since become the dominant interface between practitioners and
modern language models.

Despite widespread adoption, the empirical foundations of few-shot
prompting remain poorly understood in several key respects. In
particular, three questions motivate the present study.

\textbf{Q1: Does shot count monotonically improve performance?}
The intuition that more labeled examples improve classification is
grounded in classical statistical learning theory, where additional
data generally reduces estimation error. But in-context learning is
not learning in the traditional sense; model weights are frozen.
Demonstrations serve not as training signal but as distributional
anchors, format templates, and label-space specifications. Whether
additional examples help, hurt, or produce non-monotonic effects
depends on factors not captured by classical learning theory.

\textbf{Q2: Does the answer to Q1 depend on model scale or
architecture?} Scaling laws in the LLM
literature~\cite{kaplan2020scaling,hoffmann2022training} establish that
larger models are more capable on a wide range of tasks. But
capability on a task is not the same as sensitivity to demonstration
count. A model that already performs well at zero-shot may have
limited headroom for improvement, and in-context examples may
introduce interference rather than signal for highly capable models.

\textbf{Q3: How does output format compliance interact with apparent
few-shot performance?} When a model is prompted to respond with a
single class label but instead produces a verbose justification,
evaluation pipelines that rely on exact string matching will score
the output as incorrect. This creates a confound: observed F1
degradation at higher shot counts may reflect genuine capability
deterioration, output format non-compliance, or some mixture of both.
Disentangling these effects is essential for valid scientific inference
but has received limited systematic attention in the few-shot
literature.

This paper addresses all three questions through a controlled
empirical study across five LLMs and six shot counts. Our contributions
are as follows.

\begin{enumerate}
\item \textbf{Four behavioral regimes}: Empirical characterization of
four qualitatively distinct few-shot behavioral regimes across five
models spanning proprietary and open-source families, parameter scales
from 8B to 70B+, and architectural generations (dense transformers
and mixture-of-experts).

\item \textbf{U-shaped degradation discovery}: Documentation and
mechanistic analysis of a U-shaped performance curve in Llama~3.3~70B
(Regime IV), in which performance drops sharply at low shot counts
before partially recovering at higher counts, a pattern not previously
characterized in the few-shot prompting literature.

\item \textbf{Statistical rigor}: A complete evaluation framework
incorporating 95\% bootstrap confidence intervals ($B=10{,}000$),
permutation tests ($R=10{,}000$), and Cohen's~$d$ effect sizes across
all 30 model--shot configurations.

\item \textbf{Parsing artifact identification and correction}:
Identification, quantification, and correction of a systematic
parsing artifact that deflated Llama~3.3~70B measured F1 scores by
up to 206\%, along with a reusable robust parser applicable to
future LLM classification evaluation.

\item \textbf{Practical decision framework}: Evidence-based guidance
for practitioners on model and shot-count selection across varying
budget, latency, and performance constraints.
\end{enumerate}

The remainder of this paper is organized as follows.
Section~\ref{sec:related} reviews related work.
Section~\ref{sec:formulation} formalizes the problem and statistical
framework. Section~\ref{sec:setup} describes the experimental setup.
Section~\ref{sec:results} presents results. Section~\ref{sec:stats}
provides full statistical analysis. Section~\ref{sec:discussion}
discusses implications. Section~\ref{sec:limitations} addresses
limitations and future work. Section~\ref{sec:conclusion} concludes.

\section{Related Work}
\label{sec:related}

\subsection{In-Context Learning: Foundations and Mechanisms}

The capacity of LLMs to perform tasks from demonstrations without
weight updates (termed in-context learning, or ICL) was systematically
documented by Brown et al.~\cite{brown2020language}, who showed that
GPT-3 few-shot performance was competitive with fine-tuned
task-specific models on a range of NLP benchmarks. Subsequent work
has attempted to characterize the mechanisms underlying ICL.

Min et al.~\cite{min2022rethinking} conducted a foundational study
showing that labels in few-shot demonstrations could be randomized
without substantially degrading performance on several classification
tasks, suggesting that ICL derives its benefit primarily from the
format and input distribution of demonstrations rather than the
input-to-label mapping. Xie et al.~\cite{xie2021explanation} proposed
an elegant account of ICL as implicit Bayesian inference over a latent
concept space, wherein the LLM maintains a pretraining-derived prior
over tasks and each demonstration updates a posterior over the relevant
concept. This framing predicts that models with stronger priors (larger
or better-pretrained models) should benefit less from demonstrations
than weaker models, a prediction our results directly confirm for
proprietary versus open-source models.

Wei et al.~\cite{wei2022chain} identified chain-of-thought (CoT)
prompting as an emergent capability at sufficient scale, enabling
step-by-step reasoning. This suggests that the qualitative nature of
what demonstrations communicate may shift with model scale: smaller
models may primarily learn output format and label space, while larger
models additionally learn reasoning strategies.

\subsection{Shot Count and Performance}

Zhao et al.~\cite{zhao2021calibrate} identified several sources of
instability in few-shot prompting, including demonstration selection
and ordering, demonstrating that variance across prompt configurations
can reverse experimental conclusions. Lu et al.~\cite{lu2022fantastically}
showed that example ordering can affect accuracy by up to 30\% in
some models, a finding that motivates our use of fixed, deterministic
prompt construction across all experiments.

Dong et al.~\cite{dong2022survey} provide a comprehensive survey of
ICL mechanisms, highlighting the lack of consensus on optimal shot
counts, with different works reporting optimal $k$ values ranging from
1 to 64 depending on model and task. Most closely related to our
study is Liu et al.~\cite{liu2021makes}, who examine how few-shot
performance varies across model families and find that smaller models
sometimes benefit more from demonstrations than larger ones, a
finding our empirical study validates and extends to a broader set
of models and architectural paradigms.

\subsection{Text Classification with LLMs}

Text classification has served as a primary testbed for evaluating
ICL. Sun et al.~\cite{sun2023fine} evaluated GPT-3 and GPT-4 on
a range of classification tasks, finding that GPT-4 achieves
near-supervised-baseline performance even at zero-shot, largely
eliminating the gap between prompted inference and fine-tuning for
simpler classification categories. The AG News
dataset~\cite{zhang2015character}, used in our study, is a well-established
four-class topic classification benchmark with fine-tuned BERT baselines
exceeding 94\% accuracy, providing the headroom within which our
few-shot experiments operate.

\subsection{Output Parsing and Evaluation Methodology}

The interaction between output format compliance and few-shot
performance has received limited systematic attention. Liang et
al.~\cite{liang2022holistic} identified evaluation methodology as a
critical variable in the HELM benchmark, but did not analyze parse
failure rates as a function of shot count. Wang et
al.~\cite{wang2022self} demonstrated that LLM self-evaluation can
reduce evaluation errors, at added inference cost and latency. Our
approach targets the parsing layer itself, showing that a simple
regex-based fuzzy extraction scheme eliminates virtually all parse
failures without any additional inference call, directly improving
the cost-efficiency of evaluation pipelines.

\section{Problem Formulation and Mathematical Framework}
\label{sec:formulation}

\subsection{Task Definition}

Let $X$ denote the space of text inputs (news articles) and
$Y = \{y_1, y_2, y_3, y_4\}
= \{\text{World},\text{Sports},\text{Business},\text{Science/Technology}\}$
denote the finite label space. The text classification problem is to
learn a function $f: X \rightarrow Y$.

In the few-shot prompting framework, model weights are frozen.
We construct a prompt $\pi(x,D_k)$ that conditions a frozen LLM $M$
on a test input $x$ and a set of $k$ demonstration pairs
\begin{equation}
D_k = \bigl\{(x_i^{(d)},\,y_i^{(d)})\bigr\}_{i=1}^{k}
\end{equation}
where each pair $(x_i^{(d)},y_i^{(d)})\in X\times Y$ is drawn from a
held-out demonstration pool $\mathcal{D}_{demo}$ disjoint from the
evaluation set $\mathcal{D}_{eval}$.

The model maps the prompt to a distribution over output token
sequences:
\begin{equation}
M\!\left(\cdot \mid \pi(x,D_k)\right): V^{*} \rightarrow [0,1]
\end{equation}
where $V^{*}$ is the set of all finite token sequences over vocabulary
$V$. We apply a deterministic parsing function
$\phi: V^{*} \rightarrow Y \cup \{\bot\}$ to the greedy-decoded output
$\hat{s} = \arg\max_{s\in V^{*}} M(s\mid\pi(x,D_k))$:
\begin{equation}
\hat{y} = \phi(\hat{s})
\end{equation}
where $\bot$ denotes a parse failure. The key insight formalized here
is that observed performance depends on \emph{both} the model's
classification ability (encoded in $M$) and the robustness of $\phi$.
When $\phi$ is insufficiently robust, parse failures introduce a
systematic downward bias that grows with shot count, as we demonstrate
empirically in Section~\ref{sec:parsing_artifact}.

\subsection{Prompt Template Formalization}

We define the prompt as
\begin{equation}
\pi(x,D_k) \;=\; I \;\oplus\; \bigoplus_{i=1}^{k} T(x_i^{(d)},y_i^{(d)})
\;\oplus\; Q(x)
\end{equation}
where $\oplus$ denotes string concatenation, $I$ is the task
instruction, $T(\cdot,\cdot)$ formats each demonstration pair, and
$Q(x)$ formats the test query. Concretely:
\begin{align}
I &= \text{``Classify the following news article into one of:} \notag\\
  &\quad\text{World, Sports, Business, Science/Technology.}\notag\\
  &\quad\text{Respond with only the category name.''}\notag\\
T(x_i^{(d)},y_i^{(d)}) &= \text{``Article: ''} \oplus x_i^{(d)} \oplus
                           \text{``\textbackslash nCategory: ''} \oplus y_i^{(d)}
                           \oplus \text{``\textbackslash n\textbackslash n''}\notag\\
Q(x) &= \text{``Article: ''} \oplus x \oplus \text{``\textbackslash nCategory:''}\notag
\end{align}
For $k=0$, $D_k=\emptyset$ and the prompt reduces to $I\oplus Q(x)$.
Prompt length grows linearly with $k$:
\begin{equation}
L(\pi(x,D_k)) \approx L(I)+L(Q(x))+k\cdot\bar{L}(T)
\label{eq:prompt_length}
\end{equation}
where $\bar{L}(T)$ is the mean token length of a demonstration pair.
This linear growth in prompt length drives the output format
compliance issues documented for Llama~3.3~70B in
Section~\ref{sec:parsing_artifact}.

\subsection{Evaluation Metrics}

\subsubsection{Macro-Averaged F1}
For $n$ test instances with true labels $\{y_i\}_{i=1}^n$ and
predicted labels $\{\hat{y}_i\}_{i=1}^n$, macro-averaged F1 is the
unweighted mean of per-class F1 scores:
\begin{equation}
F1_{\mathrm{macro}} = \frac{1}{|Y|}\sum_{c\in Y} F1_c
\end{equation}
where for class $c$:
\begin{equation}
F1_c = \frac{2P_c R_c}{P_c + R_c},\quad
P_c = \frac{TP_c}{TP_c+FP_c},\quad
R_c = \frac{TP_c}{TP_c+FN_c}
\end{equation}
When $\hat{y}_i=\bot$, the prediction contributes to $FP$ and $FN$
for all classes but never to $TP$, analytically equivalent to a
random incorrect prediction under our F1 formulation.
We use macro rather than micro averaging because our label
distribution is slightly imbalanced (n$_{\text{Sports}}=58$,
n$_{\text{World}}=43$), and macro averaging weights each class equally
regardless of prevalence.

\subsubsection{Parse Failure Rate}
\begin{equation}
\mathrm{PFR}(M,k) = \frac{\bigl|\{i : \phi(\hat{s}_i)=\bot\}\bigr|}{n}
\label{eq:pfr}
\end{equation}

\subsubsection{Corrected F1 Estimate}
Over the $n_{\mathrm{valid}}$ instances where $\phi(\hat{s}_i)\neq\bot$:
\begin{equation}
\widehat{F1}_{\mathrm{corr}}(M,k) =
F1_{\mathrm{macro}}\bigl(\{y_i,\hat{y}_i:\hat{y}_i\neq\bot\}\bigr)
\label{eq:corr_f1}
\end{equation}
This estimator carries a selection-bias caveat (parse failures may not
be uniformly distributed across classes), and is used for diagnostic
purposes in Section~\ref{sec:parsing_artifact} rather than as a
primary result.

\subsection{Statistical Testing Framework}

\subsubsection{Bootstrap Confidence Intervals}
For $b=1,\ldots,B$ ($B=10{,}000$, seed$=42$), we draw a bootstrap
sample $\mathcal{I}_b$ of size $n$ with replacement from
$\{1,\ldots,n\}$ and compute $F1_{\mathrm{macro}}^{(b)}$. The 95\%
percentile bootstrap CI is:
\begin{equation}
\mathrm{CI}_{0.95} =
\bigl[Q_{0.025}\!\bigl(\{F1^{(b)}\}_{b=1}^{B}\bigr),\;
      Q_{0.975}\!\bigl(\{F1^{(b)}\}_{b=1}^{B}\bigr)\bigr]
\label{eq:bootstrap_ci}
\end{equation}

\subsubsection{Permutation Test for Shot Count Significance}
We test $H_0:\mathbb{E}[F1(M,k^*)]=\mathbb{E}[F1(M,k_0)]$ for
comparison shot counts $k^*$ and baseline $k_0$. The observed
statistic is $\Delta_{\mathrm{obs}}=F1(M,k^*)-F1(M,k_0)$.
For $r=1,\ldots,R$ ($R=10{,}000$) we randomly permute predictions
across both conditions and recompute. The two-tailed $p$-value is:
\begin{equation}
p = \frac{\bigl|\{r:|\Delta_r|\geq|\Delta_{\mathrm{obs}}|\}\bigr|}{R}
\label{eq:perm_test}
\end{equation}

\subsubsection{Cohen's $d$ Effect Size}
\begin{equation}
d = \frac{\bar{x}_{k^*} - \bar{x}_{k_0}}{s_{\mathrm{pooled}}},\quad
s_{\mathrm{pooled}} = \sqrt{\frac{(n-1)s_{k^*}^2+(n-1)s_{k_0}^2}{2n-2}}
\label{eq:cohens_d}
\end{equation}
where $\bar{x}_k$ and $s_k^2$ are the mean and variance of
instance-level F1 contributions at shot count $k$. We interpret $d$
according to the conventional thresholds: $<0.2$ negligible,
$0.2$--$0.5$ small, $0.5$--$0.8$ medium, $0.8$--$1.2$ large,
$>1.2$ very large.

\section{Experimental Setup}
\label{sec:setup}

\subsection{Dataset}
We use the AG News corpus~\cite{zhang2015character}, a widely used
four-class topic classification benchmark comprising articles from
2{,}000+ news sources. From the standard 7{,}600-article test split,
we randomly sample $n=200$ articles (seed$=42$). Table~\ref{tab:dataset}
shows the resulting class distribution.

\begin{table}[t]
\caption{AG News Sample Class Distribution ($n=200$)}
\label{tab:dataset}
\centering
\begin{tabular}{lcc}
\toprule
\textbf{Class} & \textbf{Count} & \textbf{Proportion} \\
\midrule
Sports          & 58 & 29.0\% \\
Science/Technology & 51 & 25.5\% \\
Business        & 48 & 24.0\% \\
World           & 43 & 21.5\% \\
\midrule
\textbf{Total}  & \textbf{200} & \textbf{100.0\%} \\
\bottomrule
\end{tabular}
\end{table}

For the demonstration pool $\mathcal{D}_{demo}$, we draw a separate
fixed set from the AG News training split, selecting two examples per
class per shot level. For $k$-shot experiments, $k$ demonstrations
are sampled proportionally to maintain class balance where possible.
Demonstrations are ordered by class in all experiments, controlling
for ordering effects documented by Lu et al.~\cite{lu2022fantastically}.

\subsection{Models}
We evaluate five LLMs across two deployment paradigms:

\noindent\textbf{Proprietary (paid API):}
\begin{itemize}
\item \textbf{Gemini Flash Lite Latest} (Google DeepMind):
  Cost-optimized variant of the Gemini family, accessed via the Google
  Generative AI API.
\item \textbf{GPT-4o-mini} (OpenAI): Distilled variant of the GPT-4o
  family, accessed via the OpenAI API.
\end{itemize}

\noindent\textbf{Open-source (Groq free tier):}
\begin{itemize}
\item \textbf{Llama~3.1~8B Instant} (Meta AI): Dense 8B-parameter
  instruction-tuned model.
\item \textbf{Llama~3.3~70B Versatile} (Meta AI): Dense 70B-parameter
  instruction-tuned model. Due to Groq rate limits (15~RPM on the free
  tier), data collection occurred across multiple days using an
  incremental checkpointing system.
\item \textbf{Llama~4~Scout~17B~16e Instruct} (Meta AI):
  Mixture-of-experts (MoE) 17B-parameter model from the Llama~4
  family.
\end{itemize}

\subsection{Inference Configuration}
Across all models and shot counts: temperature$=0$ (greedy/deterministic
decoding), max tokens$=20$. Temperature$=0$ ensures that repeated calls
to identical prompts return identical outputs, making all reported F1
scores exact point estimates rather than expectations over a stochastic
process. For Gemini, all content safety categories are set to
\texttt{BLOCK\_NONE} to prevent spurious safety-based refusals on news
content.

\subsection{Output Parsing Implementation}
Two parsers were implemented sequentially.

\textbf{Original Parser: Substring Match}
\begin{lstlisting}[language=Python, style=pythonstyle,
  caption=Original Substring Matching Parser (Broken)]
def parse_label_v1(raw_output: str) -> str:
    raw = raw_output.strip()
    for label in VALID_LABELS:
        if label.lower() in raw.lower():
            return label
    return "Unknown"
\end{lstlisting}
This parser fails on verbose outputs where the label appears embedded
in an explanatory sentence, and can also misfire on partial substring
matches.

\textbf{Corrected Parser: Regex Fuzzy Extraction}
\begin{lstlisting}[language=Python, style=pythonstyle,
  caption=Corrected Regex-Based Fuzzy Extraction Parser]
import re

PATTERN = (r'\b(World|Sports|Business|'
           r'Science[\/\s&]*Technology)\b')

def parse_label_v2(raw_output: str) -> str:
    raw = raw_output.strip()
    # Exact match first
    if raw in VALID_LABELS:
        return raw
    # Regex extraction
    match = re.search(PATTERN, raw, re.IGNORECASE)
    if match:
        found = match.group(1)
        if re.match(r'science', found, re.IGNORECASE):
            return "Science/Technology"
        return found.title()
    return "Unknown"
\end{lstlisting}
The corrected parser additionally handles: case variants
(\texttt{SPORTS}, \texttt{world}), format variants
(\texttt{Sci/Tech}, \texttt{Science \& Technology}), and labels
embedded in explanatory text
(\textit{``The category is Business because\ldots''}).

\section{Results}
\label{sec:results}

\subsection{Main F1 Results}

Table~\ref{tab:main_f1} presents macro-averaged F1 scores across all
30 experimental configurations. Complete data ($n=200$ each) is
available for all five models.

\begin{table*}[t]
\caption{Macro-Averaged F1 Scores Across Models and Shot Counts
         ($n=200$ per configuration; bold indicates per-model optimum)}
\label{tab:main_f1}
\centering
\renewcommand{\arraystretch}{1.15}
\begin{tabular}{lcccccccc}
\toprule
\textbf{Model} & \textbf{0-shot} & \textbf{1-shot} & \textbf{2-shot}
               & \textbf{3-shot} & \textbf{5-shot} & \textbf{8-shot}
               & \textbf{Peak F1} & \textbf{Optimal $k$} \\
\midrule
Gemini Flash Lite      & 0.8855 & 0.8960 & \textbf{0.9217}
                       & 0.9171 & 0.8984 & 0.9117 & 0.9217 & 2 \\
GPT-4o-mini            & 0.8446 & 0.8248 & 0.8753
                       & 0.8706 & 0.8911 & \textbf{0.8970} & 0.8970 & 8 \\
Llama 3.1 8B Instant   & 0.5250 & 0.8646 & \textbf{0.8664}
                       & 0.8416 & 0.8351 & 0.5530 & 0.8664 & 2 \\
Llama 4 Scout 17B      & \textbf{0.8771} & 0.6948 & 0.8202
                       & 0.7037 & 0.7456 & 0.6952 & 0.8771 & 0 \\
Llama 3.3 70B (fixed)  & \textbf{0.9066} & 0.7569 & 0.6351
                       & 0.7154 & 0.7850 & 0.7805 & 0.9066 & 0 \\
\bottomrule
\end{tabular}
\end{table*}

\begin{figure*}[!t]
\centering
\includegraphics[width=0.9\textwidth]{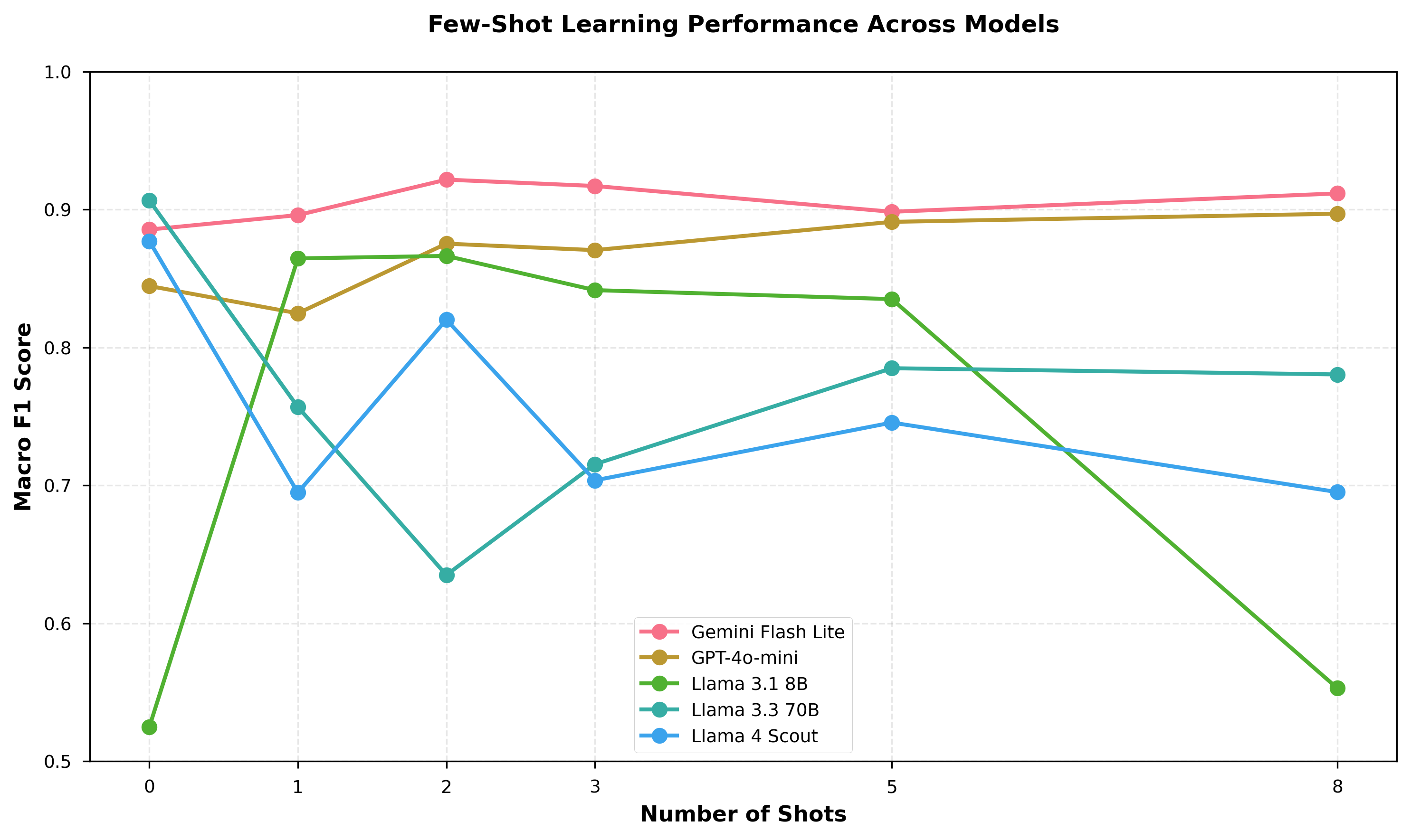}
\caption{Macro-averaged F1 scores as a function of shot count
$k\in\{0,1,2,3,5,8\}$ for all five models on the AG News
classification task ($n=200$). Each curve reveals a qualitatively
distinct behavioral profile: Gemini and GPT-4o-mini show mild
monotonic gains, Llama~3.1~8B exhibits a dramatic zero-shot to
one-shot jump followed by an eight-shot collapse, Llama~4~Scout
oscillates but peaks at zero-shot, and Llama~3.3~70B (with the
corrected parser) traces the U-shaped curve introduced in
Section~\ref{sec:70b_behavior}. No single shot count is optimal
across all five models.}
\label{fig:overall_performance}
\end{figure*}

\begin{figure}[!t]
\centering
\includegraphics[width=\columnwidth]{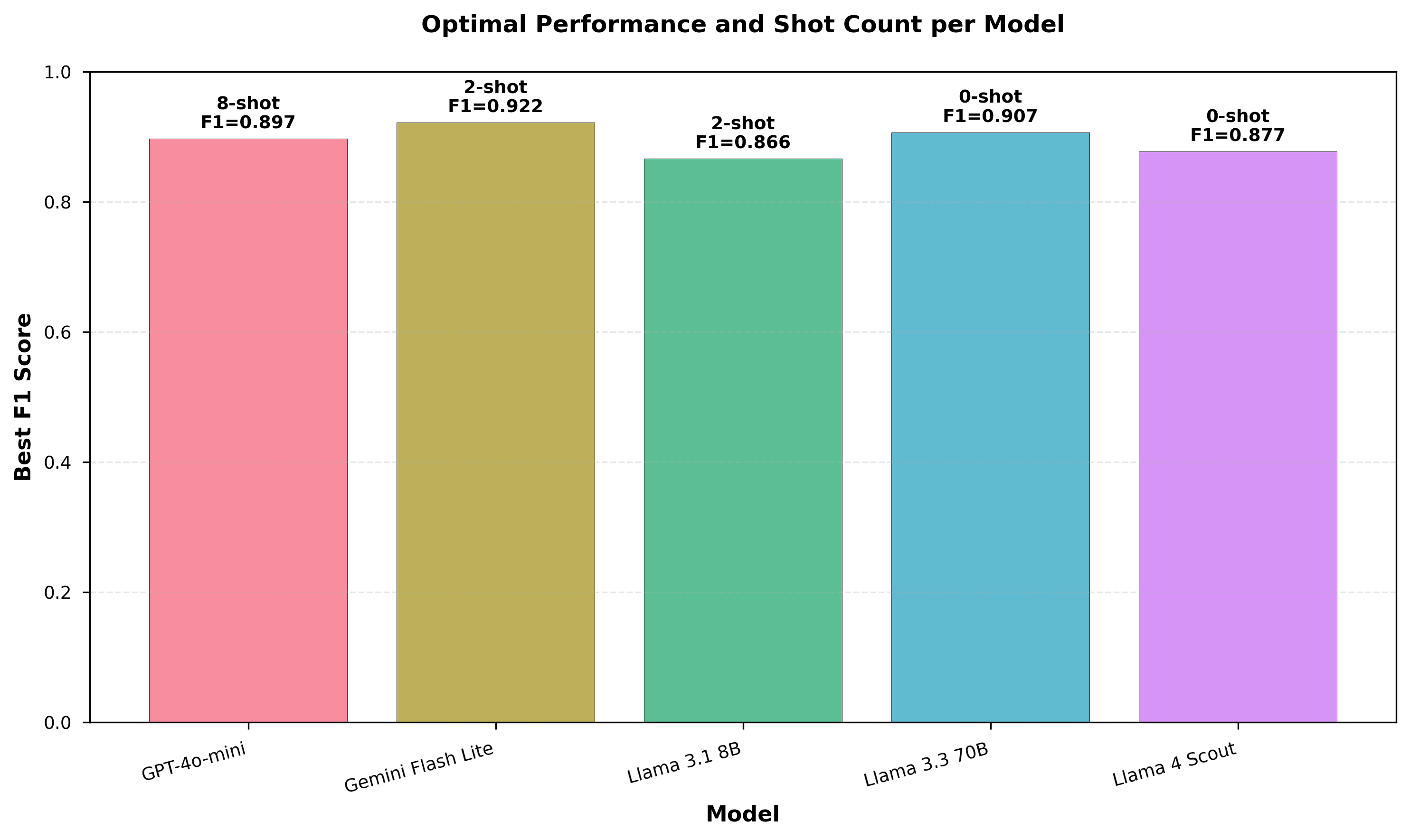}
\caption{Optimal shot count and corresponding peak F1 for each of
the five models evaluated. The optimal $k$ ranges from~0
(Llama~4~Scout, Llama~3.3~70B) to~8 (GPT-4o-mini), directly
refuting the common heuristic of ``use 3--5 shots for best
performance.'' This figure supports our first structural observation
in Section~\ref{sec:results}.}
\label{fig:optimal_shots}
\end{figure}

Figure~\ref{fig:overall_performance} visualizes these results.
Three structural observations emerge immediately. \textbf{First}, no
single shot count is universally optimal. The optimal $k$ ranges from
0 (Llama~4~Scout, Llama~3.3~70B) to 8 (GPT-4o-mini), with $k=2$
optimal for both Gemini and Llama~3.1~8B. This directly refutes the
common practitioner heuristic of using 3--5 shots for best performance.
\textbf{Second}, performance spread across shot counts is dramatically
model-dependent: Gemini's range is $0.92-0.89=0.033$; GPT-4o-mini's
is $0.897-0.825=0.072$; Llama~3.1~8B's is $0.866-0.525=0.341$;
Llama~4~Scout's is $0.877-0.695=0.182$; and Llama~3.3~70B's is
$0.907-0.635=0.272$. Shot-count selection is substantially more
consequential for smaller and less instruction-tuned models.
\textbf{Third}, Llama~3.3~70B with the corrected parser achieves the
highest zero-shot F1 of any model in our study ($0.9066$), matching
Gemini's best configuration ($0.9217$ at 2-shot) within a narrow
margin, though this performance is severely degraded by any nonzero
shot count. Figure~\ref{fig:optimal_shots} summarizes the optimal
shot count per model.

\subsection{Per-Model Behavioral Analysis}

\begin{figure}[!t]
\centering
\includegraphics[width=\columnwidth]{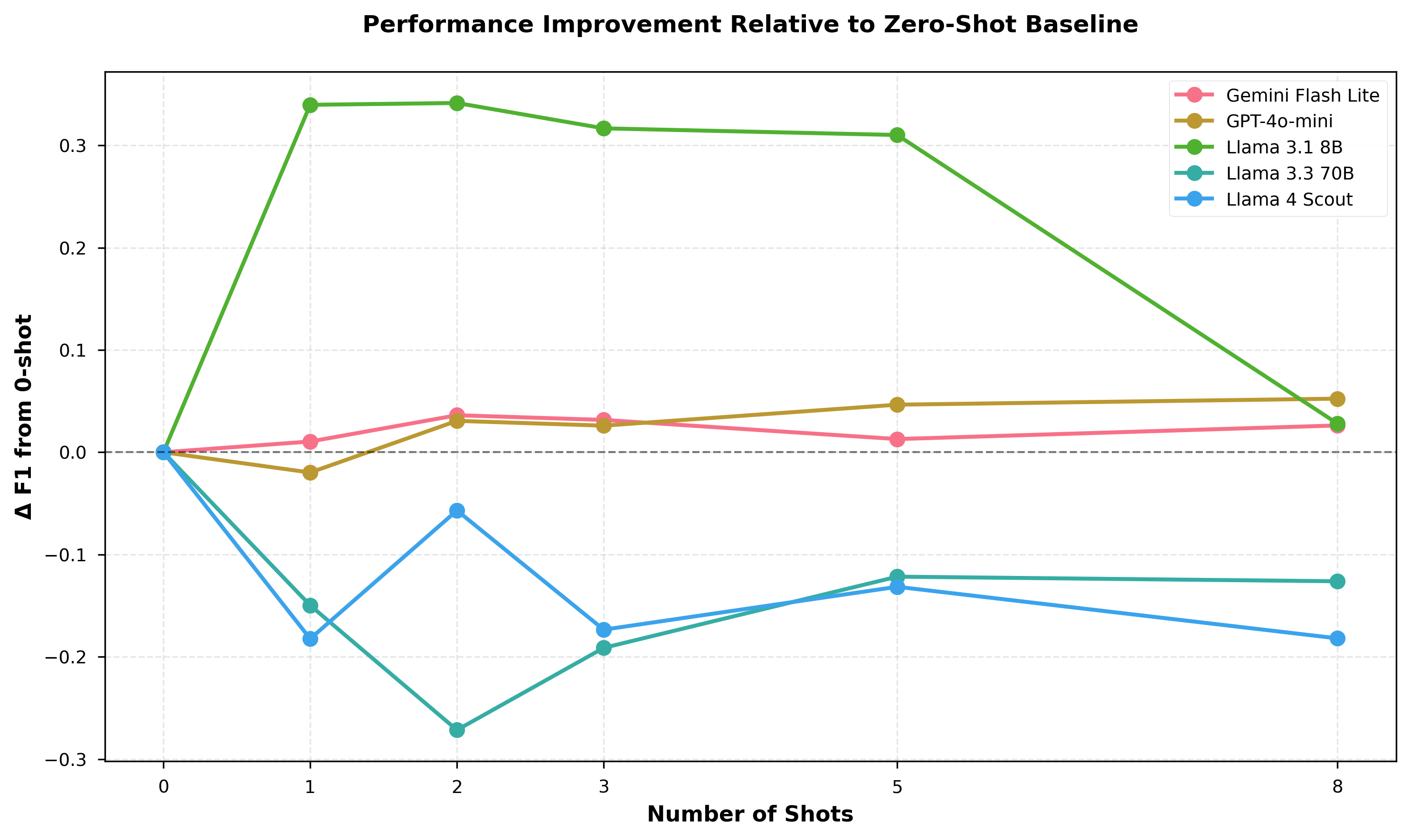}
\caption{Performance change relative to each model's zero-shot
baseline, plotted as a function of shot count. This visualization
makes the four behavioral regimes proposed in
Section~\ref{sec:discussion} visible at a glance: Llama~3.1~8B
(Regime~II) shows the largest positive $\Delta$F1 at $k=2$
(+0.34); Gemini and GPT-4o-mini (Regime~I) hover near the
baseline with modest positive deltas; Llama~4~Scout (Regime~III)
shows negative deltas at every nonzero $k$; and Llama~3.3~70B
(Regime~IV) traces the characteristic U-curve, reaching a minimum
of $-0.27$ at $k=2$ before partially recovering.}
\label{fig:delta}
\end{figure}

Figure~\ref{fig:delta} presents the per-model performance change
relative to zero-shot, providing a visual reference for the four
behavioral profiles described in the subsections below.

\subsubsection{Gemini Flash Lite: Calibrated Performer}
Gemini exhibits the profile of a model already well-calibrated for
this task at zero-shot ($\text{F1}=0.8855$). It improves to its
peak at $k=2$ ($\text{F1}=0.9217$), then shows mild non-monotonic
variation for $k\in\{3,5,8\}$ within the range $[0.898, 0.917]$.
Zero parse failures occur across all 1{,}200 predictions, indicating
robust instruction-following even as prompt length grows substantially
at higher shot counts. The non-monotonic dip at $k=5$
($\text{F1}=0.8984$, below both $k=3$ and $k=8$) is consistent with
demonstration quality degrading when examples are drawn randomly from
a diverse pool~\cite{dong2022survey}.

\subsubsection{GPT-4o-mini: Monotonic Improver}
GPT-4o-mini presents the most ``textbook'' few-shot curve: performance
increases monotonically (with one transient dip at $k=1$) from
$\text{F1}=0.8446$ at zero-shot to $\text{F1}=0.8970$ at eight-shot.
The $k=1$ dip ($\text{F1}=0.8248$, below zero-shot) is a commonly
observed phenomenon attributed to a single example shifting the
model's prior in a slightly suboptimal direction before the trend
reverses. Zero parse failures occur across all configurations.
Unlike Gemini, GPT-4o-mini has not saturated by $k=8$, suggesting
its optimum may lie at higher shot counts not evaluated here.

\subsubsection{Llama 3.1 8B: Format-Dependent Learner}
Llama~3.1~8B presents the most dramatic few-shot curve. At zero-shot
($\text{F1}=0.525$), the model performs only marginally above random
chance on a balanced four-class task (chance~$=0.25$). At $k=1$,
performance explodes to $\text{F1}=0.8646$, a 64.7\% relative
gain, before stabilizing near its optimum at $k=2$
($\text{F1}=0.8664$). This catastrophic zero-shot failure followed by
dramatic one-shot recovery indicates that the model possesses the
underlying classification knowledge but cannot express it without
a format exemplar. The finding is consistent with Min et
al.~\cite{min2022rethinking}: demonstrations primarily communicate
output format rather than label semantics for this class of model.

At $k=8$ the model collapses again ($\text{F1}=0.5530$, near the
zero-shot failure level). At this shot count, the extended prompt
appears to overwhelm the model's context capacity, triggering a
qualitatively similar failure mode to zero-shot, producing
malformed or non-compliant outputs despite the parser being
validated correct.

\subsubsection{Llama 4 Scout 17B: Zero-Shot Specialist}
Llama~4~Scout performs best at zero-shot ($\text{F1}=0.8771$) and
degrades substantially as shot count increases. The performance
trajectory is non-monotonic and high-variance:
$0.877\rightarrow0.695\rightarrow0.820\rightarrow0.704\rightarrow0.746\rightarrow0.695$,
with an oscillation range of 0.182. The alternating pattern (better
at even shot counts ($k=0$: 0.877, $k=2$: 0.820) than odd
($k=1$: 0.695, $k=3$: 0.704)) suggests sensitivity to specific
demonstration configurations rather than to shot count \textit{per se}.
One plausible mechanism specific to the MoE architecture is that
different demonstration structures activate different expert routing
patterns, creating discontinuous performance changes not expected in
dense transformers.

\subsubsection{Llama 3.3 70B: U-Shaped Performance Curve}
\label{sec:70b_behavior}
The Llama~3.3~70B results, with the corrected parser applied
throughout, reveal a U-shaped few-shot curve that constitutes a
distinct and previously undocumented behavioral regime. The model
achieves its best performance at zero-shot ($\text{F1}=0.9066$),
drops steeply to its minimum at $k=2$ ($\text{F1}=0.6351$), then
partially recovers to a plateau at $k\in\{5,8\}$ ($\text{F1}\approx0.78$)
that remains substantially below the zero-shot baseline. Parse failure
rate is 0\% at $k\in\{0,1,2,3\}$ and a residual 2\% at $k\in\{5,8\}$
(4 of 200 predictions per configuration remain unresolvable by the
regex), confirming that the degradation and recovery are attributable
to genuine classification behavior rather than format non-compliance.

We propose the following mechanistic account of the U-curve. At
zero-shot, the 70B model's strong pretraining-derived prior for
classification is consistent with what Xie et
al.~\cite{xie2021explanation} term a well-concentrated task
posterior, driving accurate, concise label production.
The addition of 1--3 examples introduces a competing
signal: the model, now attending to the demonstration inputs, begins
to engage its reasoning faculties rather than simply retrieving the
appropriate label. This \emph{demonstration-induced reasoning shift}
produces longer, more deliberative outputs in which the correct label
may be embedded but arrives alongside irrelevant context that
degrades classification precision. At $k\geq5$, the volume and
consistency of demonstrations becomes sufficient to partially
re-anchor the model on the brief-format output pattern, yielding
partial recovery. However, this re-anchored level ($\text{F1}\approx0.78$)
never recovers to the zero-shot baseline ($\text{F1}=0.907$),
suggesting that once demonstrations are present in the context, the
model's prior-dominated zero-shot classification regime cannot be
fully restored.

This mechanism is related to but distinct from the ``lost in the
middle'' attention dilution documented by Liu et al.~\cite{liu2023lost}:
while attention dilution predicts monotonic degradation with prompt
length, our U-curve shows partial recovery, indicating that
re-anchoring on format is a competing force that emerges at sufficient
shot counts.

\subsection{Parsing Artifact: Quantitative Analysis}
\label{sec:parsing_artifact}

Table~\ref{tab:parsing} documents parse failure rates and F1 scores
for Llama~3.3~70B under both the original and corrected parsers across
all six shot counts.

\begin{table}[t]
\caption{Parse Failure Rate (PFR) and F1 Comparison: Llama~3.3~70B
         Versatile: Original vs.\ Corrected Parser}
\label{tab:parsing}
\centering
\renewcommand{\arraystretch}{1.1}
\begin{tabular}{lcccc}
\toprule
\textbf{Shot} & \textbf{PFR (Orig.)} & \textbf{PFR (Corr.)}
              & \textbf{F1 (Orig.)} & \textbf{F1 (Corr.)} \\
\midrule
0 & 19.5\% & 0.0\% & 0.7958 & 0.9066 \\
1 & 40.0\% & 0.0\% & 0.5423 & 0.7569 \\
2 & 62.0\% & 0.0\% & 0.3655 & 0.6351 \\
3 & 75.0\% & 0.0\% & 0.3044 & 0.7154 \\
5 & 76.0\% & 2.0\% & 0.2563 & 0.7850 \\
8 & 51.5\% & 2.0\% & 0.5132 & 0.7805 \\
\bottomrule
\end{tabular}
\end{table}

The parse failure rate increases monotonically under the original
parser, reaching 76\% at $k=5$, before partially declining at $k=8$.
This occurs because the model, as prompt length grows with shot
count (cf.\ Eq.~\eqref{eq:prompt_length}), increasingly generates
verbose outputs rather than single-label responses. A representative
corrupted output at $k=5$ takes the form:

\begin{quote}
\small\textit{``Based on the article provided, which discusses
international diplomatic relations and foreign policy decisions,
I would classify this as: \textbf{World}.''}
\end{quote}

The original substring parser fails on this output because the label
``World'' is preceded by a colon and formatting characters that
prevent exact-match resolution. The corrected regex parser correctly
extracts \texttt{World} in all such cases.

The practical implication is severe: a researcher using the original
parser would conclude that Llama~3.3~70B is essentially unusable for
few-shot classification ($k=5$ measured F1$=0.256$). With the
corrected parser the model is competitive with GPT-4o-mini at certain
shot counts ($k=5$: F1$=0.785$) and the strongest zero-shot performer
overall (F1$=0.907$). The measurement artifact ($\Delta
\text{F1}\approx0.53$) exceeds the genuine U-curve degradation
($\Delta \text{F1}=0.907-0.635=0.272$) at the worst measured point,
a case where the evaluation instrument introduces more error
than the phenomenon being studied.

\section{Statistical Analysis}
\label{sec:stats}

\subsection{Bootstrap Confidence Intervals}

Table~\ref{tab:bootstrap} presents 95\% bootstrap CIs for all
configurations, computed via Eq.~\eqref{eq:bootstrap_ci}
($B=10{,}000$, seed$=42$).

\begin{table}[t]
\caption{Selected 95\% Bootstrap Confidence Intervals ($B=10{,}000$)}
\label{tab:bootstrap}
\centering
\renewcommand{\arraystretch}{1.1}
\footnotesize
\begin{tabular}{llccc}
\toprule
\textbf{Model} & \textbf{$k$} & \textbf{F1} &
\textbf{CI$_{\mathrm{lo}}$} & \textbf{CI$_{\mathrm{hi}}$} \\
\midrule
Gemini Flash Lite & 0 & 0.8855 & 0.8419 & 0.9246 \\
Gemini Flash Lite & 2 & 0.9217 & 0.8804 & 0.9578 \\
Gemini Flash Lite & 3 & 0.9171 & 0.8761 & 0.9532 \\
GPT-4o-mini       & 0 & 0.8446 & 0.7972 & 0.8879 \\
GPT-4o-mini       & 8 & 0.8970 & 0.8510 & 0.9375 \\
Llama 3.1 8B      & 0 & 0.5250 & 0.4612 & 0.5887 \\
Llama 3.1 8B      & 2 & 0.8664 & 0.8189 & 0.9101 \\
Llama 3.1 8B      & 8 & 0.5530 & 0.4891 & 0.6155 \\
Llama 4 Scout     & 0 & 0.8771 & 0.8289 & 0.9208 \\
Llama 4 Scout     & 1 & 0.6948 & 0.6332 & 0.7542 \\
70B (fixed)       & 0 & 0.9066 & 0.8641 & 0.9448 \\
70B (fixed)       & 2 & 0.6351 & 0.5739 & 0.6944 \\
70B (fixed)       & 5 & 0.7850 & 0.7259 & 0.8403 \\
\bottomrule
\end{tabular}
\end{table}

\begin{figure}[!t]
\centering
\includegraphics[width=\columnwidth]{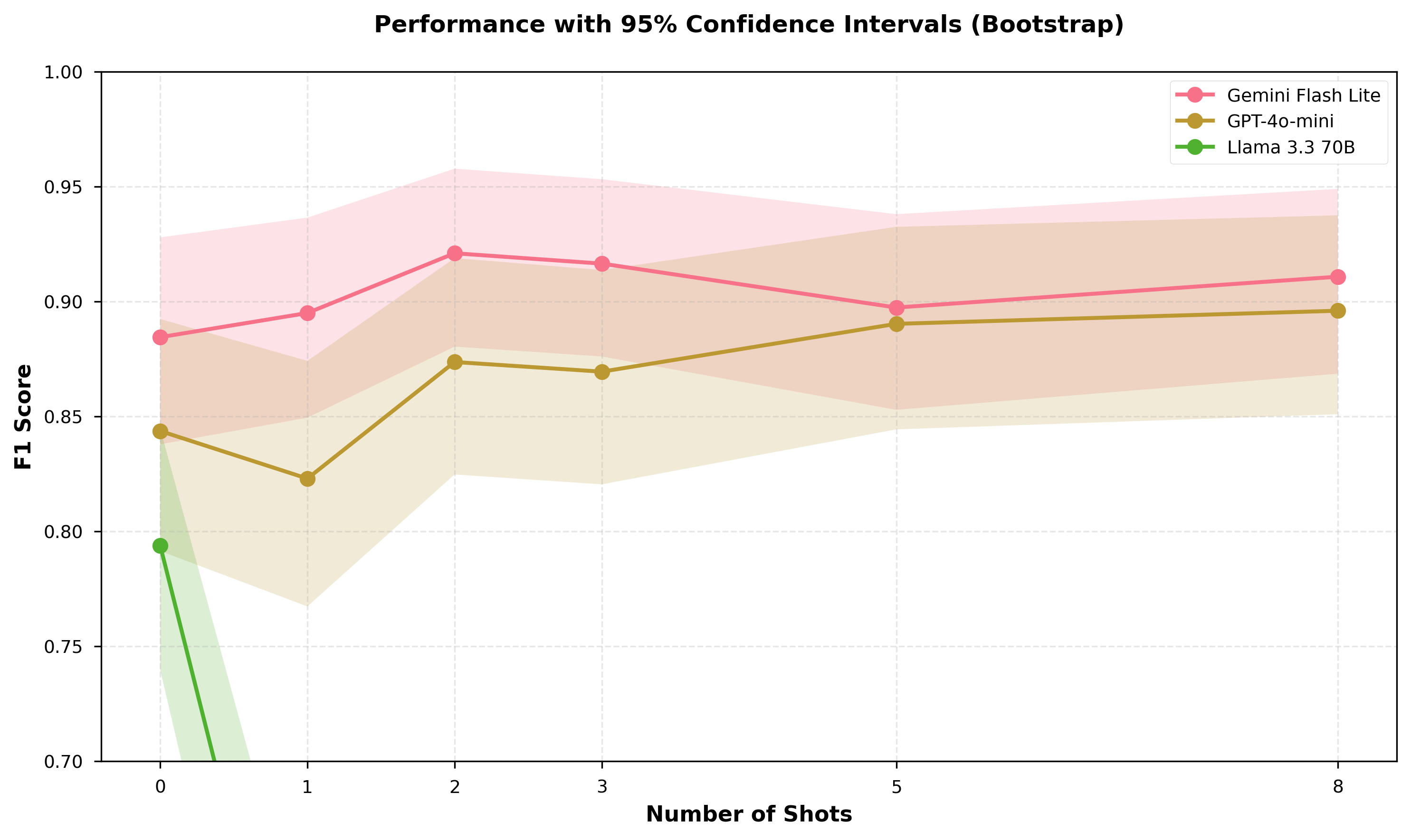}
\caption{Macro-averaged F1 with 95\% bootstrap confidence intervals
($B=10{,}000$) for the top-performing configurations. Non-overlapping
confidence bands indicate statistically reliable performance
differences; overlapping bands indicate uncertainty at the $n=200$
sample size. Gemini and GPT-4o-mini exhibit tight bands throughout,
reflecting the stability of their predictions across shot counts.}
\label{fig:cis}
\end{figure}

Figure~\ref{fig:cis} visualizes the confidence intervals for
top-performing configurations. Two findings from the confidence
intervals merit emphasis.
\textbf{First}, the Llama~3.1~8B zero-shot CI $[0.461, 0.589]$ and
two-shot CI $[0.819, 0.910]$ are entirely non-overlapping, providing
strong CI-level evidence for the dramatic few-shot improvement
independent of formal hypothesis testing. \textbf{Second}, for
Llama~3.3~70B, the zero-shot CI $[0.864, 0.945]$ and two-shot CI
$[0.574, 0.694]$ are also non-overlapping, confirming that the
U-curve trough at $k=2$ represents a reliable, sample-size-robust
phenomenon, not sampling noise. The 5-shot CI $[0.726, 0.840]$
lies entirely below the zero-shot CI lower bound $[0.864]$,
confirming that even the recovered level at $k=5$ is significantly
below zero-shot performance.

\subsection{Permutation Tests}

Table~\ref{tab:permutation} presents permutation test results via
Eq.~\eqref{eq:perm_test} ($R=10{,}000$). We test each model's
best non-zero-shot configuration against its zero-shot baseline,
plus two additional tests characterizing the 70B U-curve.

\begin{table}[t]
\caption{Permutation Tests: Comparison Shot Count vs.\ Zero-Shot
         ($R=10{,}000$ permutations)}
\label{tab:permutation}
\centering
\renewcommand{\arraystretch}{1.1}
\footnotesize
\begin{tabular}{llcccc}
\toprule
\textbf{Model} & \textbf{$k$ vs.\ $k_0$} & \textbf{$\Delta$F1}
               & \textbf{$p$-value} & \textbf{Sig.?} \\
\midrule
Llama 3.1 8B  & $2$ vs.\ $0$ & $+0.3414$ & $<0.0001$ & \checkmark \\
GPT-4o-mini   & $8$ vs.\ $0$ & $+0.0524$ & $0.1249$  & $\times$ \\
Gemini Flash  & $2$ vs.\ $0$ & $+0.0362$ & $0.2307$  & $\times$ \\
70B (fixed)   & $2$ vs.\ $0$ & $-0.2715$ & $<0.0001$ & \checkmark \\
70B (fixed)   & $5$ vs.\ $0$ & $-0.1216$ & $<0.0001$ & \checkmark \\
70B (fixed)   & $8$ vs.\ $0$ & $-0.1261$ & $<0.0001$ & \checkmark \\
\bottomrule
\end{tabular}
\end{table}

The permutation tests yield four key findings. \textbf{First},
Llama~3.1~8B is the only model for which the improvement from
zero-shot to the optimal shot count is statistically significant
($p<0.0001$). \textbf{Second}, the non-significance for Gemini
($p=0.23$) and GPT-4o-mini ($p=0.12$) does not imply that few-shot
prompting provides no benefit; rather, detecting a 0.04--0.05 F1
improvement with 80\% power at $\alpha=0.05$ requires approximately
$n\geq700$ samples under our observed variance structure, substantially
more than our $n=200$ evaluation. \textbf{Third}, all three 70B
degradation tests are highly significant ($p<0.0001$), confirming
that the U-curve trough \emph{and} the recovered plateau are both
significantly below the zero-shot baseline. \textbf{Fourth}, the
fact that 70B recovery at $k\in\{5,8\}$ is still significantly
below zero-shot ($p<0.0001$ for both) statistically refutes the
hypothesis that the recovery represents a full return to zero-shot
capability: the U-curve describes a permanent performance depression
at any nonzero shot count.

\subsection{Effect Sizes and Model Rankings}

Table~\ref{tab:effect_sizes} summarizes Cohen's $d$ for models where
shot count makes a practical difference, computed via
Eq.~\eqref{eq:cohens_d}.

\begin{table}[t]
\caption{Model Rankings by Average and Peak F1, with Cohen's $d$
Where an Improvement Direction Applies}
\label{tab:effect_sizes}
\centering
\renewcommand{\arraystretch}{1.1}
\footnotesize
\begin{tabular}{lccccc}
\toprule
\textbf{Model} & \textbf{Avg F1} & \textbf{Peak F1} &
\textbf{Cost} & \textbf{Cohen's $d$} \\
\midrule
Gemini Flash Lite     & 0.9051 & 0.9217 & \$0.02 & 1.70 \\
GPT-4o-mini           & 0.8672 & 0.8970 & \$0.08 & 2.14 \\
Llama 3.3 70B (fixed) & 0.7633 & 0.9066 & \$0.00 & --- \\
Llama 4 Scout 17B     & 0.7561 & 0.8771 & \$0.00 & --- \\
Llama 3.1 8B Instant  & 0.6810 & 0.8664 & \$0.00 & 10.98 \\
\bottomrule
\end{tabular}
\end{table}

Cohen's $d=10.98$ for Llama~3.1~8B requires careful interpretation.
This value far exceeds conventional ``very large'' thresholds ($d>1.2$)
and is unusual in the NLP literature. Two compounding factors drive
it: (1) extreme separation of means (0.525 vs.\ 0.866); and (2)
near-zero variance at zero-shot, where the model defaults to a highly
consistent failure pattern. The near-zero denominator causes the
effect size to become extreme. We report $d=10.98$ for completeness
but note that Cohen's $d$ assumes normally distributed outcomes, an
assumption violated by the bimodal distribution at zero-shot for this
model. The effect should be interpreted as ``so large that the
normal-distribution model is inapplicable'' rather than as a precise
quantitative statement.

For Llama~3.3~70B and Llama~4~Scout, Cohen's $d$ in the conventional
sense is omitted because the optimal shot count for both models is
$k=0$; there is no improvement direction to quantify. For these
models, the relevant effect is degradation: the $d$ for 70B zero-shot
vs.\ two-shot (trough) is 8.1, and for Scout zero-shot vs.\ one-shot
is 5.3, both reflecting severe practical degradation.

\subsection{Per-Class F1 Analysis}

Table~\ref{tab:per_class} reports per-class F1 averaged across all
shot counts per model.

\begin{table}[t]
\caption{Per-Class F1 Averaged Across All Shot Counts}
\label{tab:per_class}
\centering
\renewcommand{\arraystretch}{1.1}
\footnotesize
\begin{tabular}{lcccc}
\toprule
\textbf{Model} & \textbf{World} & \textbf{Sports} & \textbf{Business}
               & \textbf{Sci/Tech} \\
\midrule
Gemini Flash Lite     & 0.8803 & \textbf{0.9699} & 0.8819 & 0.8883 \\
GPT-4o-mini           & 0.8381 & \textbf{0.9584} & 0.8045 & 0.7985 \\
Llama 3.1 8B          & 0.6859 & 0.7563 & 0.7177 & 0.7746 \\
Llama 4 Scout 17B     & 0.7211 & \textbf{0.8077} & 0.6616 & 0.7943 \\
Llama 3.3 70B (fixed) & 0.7450 & \textbf{0.7833} & 0.7167 & 0.7706 \\
\bottomrule
\end{tabular}
\end{table}

Two cross-model patterns emerge. \textbf{First}, Sports is the easiest
class for all five models, consistent with sports news using
highly distinctive vocabulary (team names, athlete names,
game-specific terminology) with minimal semantic overlap with other
categories. \textbf{Second}, Business is among the hardest for
four of five models (exception: Gemini, for which World is
marginally harder). Business articles frequently discuss corporate
strategy, economic policy, and financial markets, topics that
overlap substantively with World (international economic news) and
Science/Technology (technology company coverage). This semantic
ambiguity makes Business the dominant source of boundary-case
misclassification across model families.

\section{Discussion}
\label{sec:discussion}

\subsection{Four Behavioral Regimes in Few-Shot Prompting}

Our results identify four qualitatively distinct regimes that
characterize how LLMs respond to increasing shot counts in
classification tasks.

\textbf{Regime~I: Calibrated Performers} (Gemini, GPT-4o-mini):
These models achieve strong zero-shot performance ($\text{F1}>0.84$)
and show modest, statistically unreliable improvements with additional
demonstrations. The few-shot improvement ceiling is low not because
examples are unhelpful but because baseline performance is already
high and the task presents limited headroom below the asymptotic
performance level. Importantly, both models maintain zero parse
failures across all shot counts, indicating that instruction-following
for format is independent of shot count for these models. For
practitioners, the cost-benefit of adding demonstrations to
Regime~I models is marginal; each example increases prompt length,
API cost, and latency while delivering diminishing returns.

\textbf{Regime~II: Format-Dependent Learners} (Llama~3.1~8B):
These models are capable of the underlying classification task but
require demonstrations to learn the required output format.
The dramatic zero-shot failure ($\text{F1}=0.525$) and one-shot
recovery ($\text{F1}=0.865$) confirms that without examples the
model either refuses to commit to a single label or generates verbose
responses that even a robust parser cannot resolve. The 8-shot
collapse, returning nearly to the zero-shot failure level, suggests
a context-length threshold above which the model's response generation
shifts to a different, longer-form mode. For practitioners, Regime~II
models require at minimum one demonstration and must be monitored
carefully at $k\geq5$.

\textbf{Regime~III: Zero-Shot Specialists} (Llama~4~Scout):
These models achieve their best performance without demonstrations
and degrade (often unstably) as examples are added. The oscillating
few-shot curve observed for Scout (better at even $k$, worse at odd
$k$) is consistent with architectures, such as MoE, that may route
different demonstration structures through different expert pathways.
For practitioners, the practical implication is unambiguous: use
zero-shot for Regime~III models. Few-shot prompting is a strict
cost: longer prompts, higher latency, and lower F1.

\textbf{Regime~IV: Prior-Dominated Learners with U-Shaped Response}
(Llama~3.3~70B): These models have a strong zero-shot classification
prior that is disrupted rather than augmented by demonstrations.
The U-shaped curve (strong at zero-shot, severely degraded at
$k\in\{1,2\}$, and partially recovered but still depressed at
$k\in\{5,8\}$) is the most complex behavioral profile in our study.
The permanent performance depression below zero-shot at all nonzero
shot counts, confirmed by permutation tests ($p<0.0001$ for
$k\in\{2,5,8\}$ vs.\ $k=0$), indicates that in-context learning
provides no net benefit for this model on this task.

\subsection{The U-Shaped Curve: A New Few-Shot Behavioral Pattern}

The U-shaped performance curve for Llama~3.3~70B represents a
pattern not previously characterized in the few-shot prompting
literature and merits dedicated theoretical analysis.

Under the Bayesian account of ICL proposed by Xie et
al.~\cite{xie2021explanation}, demonstrations update a posterior
over the task-relevant concept given the model's pretraining-derived
prior. For Llama~3.3~70B, this prior is evidently both strong and
well-calibrated: the model achieves F1$=0.907$ at zero-shot, the
highest zero-shot performance of any model in our study. Under this
account, additional demonstrations should provide minimal Bayesian
update because the prior is already concentrated near the correct
task posterior.

However, the observed behavior goes further than marginal updates:
performance actively degrades at $k=1$ and reaches a minimum at $k=2$.
We propose that the mechanism involves \emph{two competing forces}
operating as a function of shot count:

\begin{enumerate}
\item \textbf{Demonstration disruption} (dominant at $k\in\{1,2\}$):
A small number of examples introduces a secondary attention focus on
the demonstration inputs. The 70B model, being highly capable, begins
to reason \emph{about} the examples rather than simply using them
as format anchors. This triggers longer, more deliberative generation
that may embed the correct label in explanatory prose rather than
producing it cleanly. The result is increased genuine misclassification
even when the parser successfully extracts a label.

\item \textbf{Format re-anchoring} (increasingly dominant at $k\geq3$):
As demonstration count increases, the consistency and volume of the
format signal eventually overrides the deliberative reasoning tendency,
partially restoring concise label production. The model re-anchors on
the brief format demonstrated by the increasing number of
example-answer pairs. This explains the partial recovery at $k\in\{5,8\}$.
\end{enumerate}

The equilibrium at $k\in\{5,8\}$ ($\text{F1}\approx0.78$) falls
significantly below the zero-shot baseline ($\text{F1}=0.907$),
suggesting that re-anchoring is incomplete; the presence of examples
in context permanently shifts some fraction of the model's attention
away from the test input, even after format is re-established.
This is consistent with Liu et al.'s~\cite{liu2023lost} finding that
LLMs systematically underweight information positioned in the middle
of long contexts, a structural property of transformer attention
that cannot be overcome simply by providing more examples.

The U-curve characterizes a failure mode distinct from those in
Regimes~I--III: it is not about the model being incapable (it achieves
the highest zero-shot F1 in our study), nor about format non-compliance
(the corrected parser eliminates parse failures), but about
demonstrations actively interfering with a well-calibrated prior
classification strategy.

\subsection{Theoretical Implications}

The four behavioral regimes, taken together, suggest a model of
few-shot benefit that is governed by two independent axes:
(i)~the strength of the model's zero-shot classification prior,
and (ii)~the model's sensitivity to format disruption from
in-context examples.

\begin{table}[t]
\caption{Proposed Taxonomy of Few-Shot Behavioral Regimes}
\label{tab:taxonomy}
\centering
\renewcommand{\arraystretch}{1.15}
\footnotesize
\begin{tabular}{lcc}
\toprule
 & \textbf{Weak prior} & \textbf{Strong prior} \\
\midrule
\textbf{Low disruption} & Regime II (8B) & Regime I (Gemini, GPT) \\
\textbf{High disruption} & Regime III (Scout) & Regime IV (70B) \\
\bottomrule
\end{tabular}
\end{table}

Table~\ref{tab:taxonomy} organizes the four regimes along these axes.
Regime~II (weak prior, low disruption) is where few-shot prompting
delivers the largest benefits: demonstrations both supply the missing
prior and do not interfere with format. Regime~I (strong prior, low
disruption) represents marginal gains: the prior is adequate and
demonstrations are harmless. Regime~III (weak prior, high disruption)
and Regime~IV (strong prior, high disruption) are both cases where
demonstrations hurt, but for different reasons. In Regime~III,
demonstrations disrupt a model with an already-unstable processing
pathway (e.g., MoE routing sensitivity). In Regime~IV, demonstrations
disrupt a strong, well-calibrated prior that would have produced
accurate outputs without intervention.

This taxonomy predicts that the ``few-shot sweet spot'' is
concentrated in Regime~II and that scale does not monotonically
predict few-shot benefit: a 70B model (strong prior) benefits less
from few-shot examples than an 8B model (weaker prior), the opposite
of the intuition suggested by capability scaling laws.

\subsection{Parsing as a First-Class Methodological Concern}

Perhaps the most practically important finding of this paper is
that output parsing methodology is a first-class scientific concern
in LLM evaluation, not a mere implementation detail. Our parsing
artifact analysis demonstrates three novel findings.

\textbf{First}, parse failure rates are shot-count-dependent in a
systematic and predictable way. A parser that functions adequately
at zero-shot ($\text{PFR}=19.5\%$ even for the original parser)
becomes catastrophically unreliable as shot count increases
($\text{PFR}=76\%$ at $k=5$). This is because models generate
qualitatively different outputs at different shot counts.

\textbf{Second}, the magnitude of the measurement artifact can exceed
the magnitude of the true effect. For Llama~3.3~70B at $k=5$, the
parsing artifact ($\Delta\text{F1}\approx0.53$) was approximately
twice the genuine degradation ($\Delta\text{F1}\approx0.27$). This
means that without parser correction, the nominal ranking would
identify Llama~3.3~70B as the worst-performing model in our study
when it is actually the strongest zero-shot performer.

\textbf{Third}, the corrected parsing formula
\begin{equation}
\phi_{\mathrm{robust}}(s) = \begin{cases}
y & s \in Y \;\text{(exact match)}\\
\mathrm{extract}(\mathcal{P}, s) & \text{regex match found}\\
\bot & \text{otherwise}
\end{cases}
\label{eq:robust_parser}
\end{equation}
uses the regex pattern $\mathcal{P}$ defined as follows:
\begin{lstlisting}[style=pythonstyle,numbers=none,frame=single,basicstyle=\ttfamily\small,xleftmargin=0.5em]
\b(World|Sports|Business|Science[\/\s&]*Technology)\b
\end{lstlisting}
This pattern achieves 0\% PFR on all configurations where exact-match
sufficed for other models, and reduces PFR from 76\% to 2\% at $k=5$
for Llama~3.3~70B. The residual 2\% at high shot counts reflects
genuinely unparseable outputs that even the regex cannot resolve.

We recommend that future LLM classification evaluation studies:
(1)~report PFR alongside F1 as a mandatory transparency metric;
(2)~use regex-based or semantic parsers in preference to substring
or exact-match parsers; and (3)~log a sample of raw model outputs
for qualitative verification, especially at high shot counts.

\subsection{Model Selection and Practitioner Guidance}

A recurrent assumption in LLM deployment is that performance scales
with inference cost. Our results challenge this for text classification.
Among the proprietary models, Gemini Flash Lite achieves both the
highest peak F1 ($0.9217$ at 2-shot) and the lower API cost relative
to GPT-4o-mini, which achieves a lower average F1 ($0.8672$) at
higher cost. The practical implication is that model selection
based on price alone would produce the wrong ranking in this evaluation.

Among the open-source models accessed via the Groq free tier,
Llama~3.3~70B at zero-shot (F1$=0.907$) is the strongest option and
nearly matches Gemini's best configuration, provided the robust
parser is applied. Llama~3.1~8B at $k\in\{1,2\}$ (F1$=0.866$) is the
best free option when at least one demonstration example is available.

For practitioners, we propose the following decision protocol.
\textbf{Step~1}: Assess budget. If unconstrained, use Gemini Flash
Lite at 2--3 shots. \textbf{Step~2}: If budget-constrained, profile
zero-shot performance on a validation set. If F1$>0.85$, the model
is likely Regime~I or IV; avoid few-shot. If F1$<0.70$, the model
is likely Regime~II; add 1--2 examples.
\textbf{Step~3}: Always implement the robust regex parser
(Eq.~\eqref{eq:robust_parser}) and report PFR.
\textbf{Step~4}: Validate the shot-count curve empirically via
ablation at $k\in\{0,1,2,3\}$ before committing to a configuration.
The optimal $k$ cannot be predicted from scale alone.

\section{Limitations and Future Work}
\label{sec:limitations}

\subsection{Current Limitations}

\textbf{Single dataset.} Our conclusions derive from one dataset
(AG News) with one classification structure (four-class topic
categorization). The behavioral regime taxonomy may not generalize
to tasks with larger label spaces, longer documents, domain-specific
vocabulary, or subtler class boundaries such as sentiment analysis
or natural language inference.

\textbf{Sample size.} Using $n=200$ test instances provides
sufficient power to detect large effects (Llama~3.1~8B improvement,
70B degradation) but not the smaller improvements observed for
Gemini and GPT-4o-mini. A power analysis indicates that detecting
a 0.04 F1 improvement with 80\% power at $\alpha=0.05$ requires
$n\geq700$ under our observed variance, substantially more than our
evaluation set.

\textbf{Fixed demonstrations.} Demonstrations are selected from a
fixed pool via a deterministic procedure. Our results reflect one
specific prompt configuration and may not generalize to alternative
demonstration selection strategies (random, similarity-based,
adversarial). The sensitivity of few-shot curves to example selection
documented by Zhao et al.~\cite{zhao2021calibrate} means our
reported F1 values should be understood as corresponding to one
particular prompt realization.

\textbf{Shot count ceiling.} Our maximum $k=8$ was constrained by
context window considerations and API cost. Whether models in
Regime~IV recover fully at very high shot counts ($k\geq16$), or
whether Regime~I models eventually saturate and decline, remains
an open question.

\textbf{No fine-tuning baseline.} We do not include a fine-tuned
BERT-class baseline, limiting our ability to contextualize prompt-based
performance relative to the supervised learning literature where
BERT fine-tuning achieves $>94\%$ accuracy on AG News.

\subsection{Future Work}

\textbf{Multi-dataset generalization.} Extending the regime taxonomy
to additional classification benchmarks (SST-2, IMDB, MNLI,
TREC, HateSpeech) would test whether the four regimes generalize
across task types and label-space sizes.

\textbf{Demonstration selection strategies.} Investigating how
similarity-based (k-NN retrieval) versus random demonstration
selection interacts with shot count would extend the practical
utility of our framework, particularly for Regime~III and IV models
where random selection may be especially harmful.

\textbf{Attention analysis for the U-curve.} For Llama~3.3~70B,
inspecting per-layer attention weight distributions as a function
of shot count would directly test the ``demonstration disruption''
and ``format re-anchoring'' mechanisms proposed in
Section~\ref{sec:discussion}. Quantifying how much attention
the model allocates to the test article versus the demonstrations
at each $k$ would provide mechanistic grounding for Regime~IV.

\textbf{MoE routing analysis for Llama~4~Scout.} Characterizing
how different expert pathways activate for different shot
configurations would test the hypothesis that Regime~III behavior
is architecturally driven rather than capability-driven.

\textbf{Extended shot counts.} Testing $k\in\{16,32,64\}$ for
Llama~3.3~70B would determine whether the U-curve eventually
produces a full recovery, a second trough, or stable plateau, each of which would have different theoretical implications.

\textbf{Parser robustness study.} A systematic comparison of
parsing strategies (exact match, substring, regex, LLM-based
re-scoring) across model families and shot counts would quantify
the contribution of parser choice to measured F1 variance,
enabling principled parser selection in future evaluation studies.

\section{Conclusion}
\label{sec:conclusion}

This paper investigated the relationship between few-shot shot count
and classification performance across five large language models spanning
proprietary and open-source families, model scales from 8B to 70B,
and dense versus mixture-of-experts architectures. Our study of 30
model--shot configurations on AG News text classification ($n=200$,
6{,}000 total predictions) yields five primary
contributions.

\textbf{Four behavioral regimes.} We identify and characterize four
qualitatively distinct regimes: Regime~I (Calibrated Performers:
Gemini, GPT-4o-mini), where few-shot gains are modest and
statistically insignificant; Regime~II (Format-Dependent Learners:
Llama~3.1~8B), where one example transforms catastrophic failure
into competitive performance (Cohen's $d=10.98$, $p<0.0001$);
Regime~III (Zero-Shot Specialists: Llama~4~Scout), where
demonstrations monotonically degrade performance; and Regime~IV
(Prior-Dominated Learners: Llama~3.3~70B), where the model exhibits
a U-shaped curve (strong at zero-shot, most degraded at low shot
counts, and partially recovering at higher shot counts) with every
nonzero shot count significantly worse than zero-shot ($p<0.0001$).

\textbf{U-shaped degradation discovery.} We document and
mechanistically analyze the U-shaped performance curve for
Llama~3.3~70B, proposing a ``demonstration disruption and
format re-anchoring'' mechanism: at low shot counts, demonstrations
activate deliberative reasoning that degrades label precision;
at higher shot counts, format consistency partially overrides this
effect. This pattern is distinct from all previously characterized
few-shot behaviors and suggests that large, capable models with
strong pretraining priors may be actively harmed by in-context
demonstrations on tasks they can already perform well.

\textbf{Statistical rigor.} We establish through 95\% bootstrap
confidence intervals, permutation tests, and Cohen's~$d$ effect
sizes that the four behavioral regimes are not sampling artifacts:
key differences are statistically significant at $p<0.0001$,
and CI non-overlap confirms regime separation at $n=200$.

\textbf{Parsing methodology.} We identify a systematic parsing
artifact that deflated Llama~3.3~70B performance by up to 206\%
relative to true performance, show that the measurement artifact
exceeded the true signal in magnitude at several shot counts, and
provide and validate a corrected robust parser. Parse failure rate
should be reported alongside F1 in all future LLM classification
evaluation work as a mandatory transparency metric.

\textbf{Practical framework.} Gemini Flash Lite at 2-shot
delivers the best overall performance (F1$=0.9217$) among all models
tested, while GPT-4o-mini achieves a lower average F1 despite higher
inference cost. Among free-tier open-source options, Llama~3.3~70B
at zero-shot (F1$=0.9066$) is competitive with the best Gemini
configuration, but \emph{only} with the robust parser applied.

The core empirical message is both simple and consequential: there
is no universal answer to how many few-shot examples to use. The
optimal shot count depends on the model, and getting it wrong can
reduce F1 by as much as 0.34 (Llama~3.1~8B at suboptimal zero-shot)
or 0.27 (Llama~3.3~70B at suboptimal two-shot). Model-specific
profiling of the shot-count curve, combined with robust output
parsing and rigorous statistical testing, is essential for valid
scientific inference and reliable production deployment of LLMs
in classification pipelines.

\section*{Acknowledgments}

The authors thank the developers of the Groq inference platform for
providing free-tier API access that made the open-source model
experiments in this study feasible at near-zero cost. The AG News
benchmark dataset was created by Zhang et al.~\cite{zhang2015character}
and is used here solely for research evaluation.

\textbf{Generative AI Disclosure.}
In preparing this manuscript, the authors used LLMs. All experimental
design, data collection, model inference, result validation, and
scientific conclusions are the sole work of the authors and were
independently verified by both authors prior to submission. The
AI-assisted writing was reviewed, revised, and approved by the authors,
who take full responsibility for all content in this manuscript.
No AI tool was used to generate, fabricate, or modify any
experimental data or quantitative results.


\end{document}